\documentclass[journal,twocolumn,final]{IEEEtran}
\usepackage[cmex10]{amsmath}
\usepackage{amssymb}
\usepackage{slashbox,pict2e}
\usepackage{multirow}
\usepackage{array}
\usepackage{mdwmath}
\usepackage{mdwtab}
\usepackage{eqparbox}
\usepackage{cite}
\usepackage{setspace,citesort}
\usepackage{pgf,pgfarrows,pgfnodes,pgfautomata,pgfheaps,pgfshade}
\usepackage{tikz,pgflibraryshapes,pgflibrarysnakes}
\usepackage{graphicx}
\usetikzlibrary{positioning,patterns,fadings}
\usetikzlibrary{decorations,decorations.pathmorphing,decorations.pathreplacing}
\usepackage{array,txfonts}
\usepackage{tikz}
\usepackage{pgffor}
\usepackage{gensymb}
\usepgflibrary{plotmarks}
\usetikzlibrary{decorations.pathreplacing} 
\usepackage{subfig}
\newcommand{\bE}{{\bf E}}
\newcommand{\bH}{{\bf H}}
\newcommand{\bJ}{{\bf J}}
\newcommand{\bj}{{\bf j}}
\newcommand{\br}{{\bf r}}

\newcommand{\grad}{\nabla}
\newcommand{\abs}[1]{\lvert#1\rvert}
\newcommand{\norm}[1]{\lVert#1\rVert}
\newcommand{\ceil}[1]{\left\lceil#1\right\rceil}
\newcommand{\Div}{\nabla\cdot}
\newcommand{\Divp}{\nabla'\cdot}
\newcommand{\curl}{\nabla\times}
\newcommand{\dt}{{\Delta t}}
\newcommand{\dow}{{\partial}}
\newcommand{\dowt}{{\partial_t}}
\newcommand{\dowtp}{{\partial_{t'}}}
\newcommand{\dowtptwo}{{\partial^2_{t'}}}
\newcommand{\dOmega}{{\partial\Omega}}
\newcommand{\dOmegam}{{{\partial\Omega}_m}}
\newcommand{\dOmegan}{{{\partial\Omega}_n}}
\newcommand{\ncross}{{\hat{n}\times}}
\newcommand{\nhat}{\hat{n}}

\providecommand{\bml}{\begin{subequations}}
\providecommand{\eml}{\end{subequations}}
\newcommand{\inprod}[2]{\langle #1, #2\rangle}
\newcommand{\inc}{\text{inc}}
\newcommand{\scat}{\text{s}}
\newcommand{\tot}{\text{t}}
\newcommand{\intr}{\text{i}}
\newcommand{\RCS}{\text{RCS}}
\newcommand{\ext}{\text{e}}
\newcommand{\fmax}{f_{\text{max}}}
\newcommand{\df}{\Delta f}
\newcommand{\ksamp}{\text{k}_{\text{samp}}}
\newcommand{\tdie}{\text{tdie}}
\newcommand{\fdie}{\text{fdie}}
\newcommand{\spctime}{(\br,t)}
\newcommand{\Etheta}{{E^{\theta}}}
\newcommand{\Ephi}{{E^{\phi}}}
\newcommand{\thetahat}{{\hat{\theta}}}
\newcommand{\phihat}{{\hat{\phi}}}
\newcommand{\zetan}{\hat{\zeta}}
\newcommand{\alphaq}{\alpha_{mn}^q}
\newcommand{\phiq}{\phi_{mn}^q}
\newcommand{\phizero}{\phi_{mn}^0}
\newcommand{\kappaq}{\kappa_{mn}^q}

\newcommand{\eps}{\varepsilon}

\newcommand{\blambda}{{\boldsymbol\lambda}}
\newcommand{\Nh}{{\text{N}_{\text{h}}}}
\newcommand{\Ns}{{\text{N}_{\text{s}}}}
\newcommand{\Nt}{{\text{N}_{\text{t}}}}
\newcommand{\p}{\text{p}}
\newcommand{\ztzs}{{z_tz_s}}

\begin{document}
\title{A Stable Higher Order Space-Time Galerkin Scheme for Time Domain Integral Equations}

\author{{A.~J.~Pray,~\IEEEmembership{Student Member,~IEEE,} ~Y. Beghein,~\IEEEmembership{Member,~IEEE,} ~N.~V.~Nair,~\IEEEmembership{Member,~IEEE, } ~K.~Cools,~\IEEEmembership{Member,~IEEE,} ~H.~Ba\u{g}c{\i}}~\IEEEmembership{Member,~IEEE,} ~and~B.~Shanker,~\IEEEmembership{Fellow,~IEEE}
\thanks{A.~J.~Pray, N.~V.~Nair, and ~B.~Shanker are with the Department of Electrical and Computer Engineering, Michigan State University, East Lansing, MI, USA, 48824 e-mail: prayandr@msu.edu .}
\thanks{Y.~Beghein is with the Department of Information Technology (INTEC), Ghent University, Belgium}
\thanks{K. Cools is with the Electrical Systems and Optics Division, University of Nottingham, Nottingham NG7 2RD, UK}
\thanks{H. Ba\u{g}c{\i} is with the Division of Computer, Electrical, and Mathematical Sciences and Engineering, King Abdullah University of Science and Technology (KAUST), 4700 KAUST, Thuwal 23955-6900, Kingdom of Saudi Arabia}
}

\maketitle

\begin{abstract}
Stability of time domain integral equation (TDIE) solvers has remained an elusive goal for many years.  Advancement of this research has largely progressed on four fronts: (1) Exact integration, (2) Lubich quadrature, (3) smooth temporal basis functions, and (4) Space-time separation of convolutions with the retarded potential.  The latter method was explored in \cite{Pray2012}.  This method's efficacy in stabilizing solutions to the time domain electric field integral equation (TD-EFIE) was demonstrated on first order surface descriptions (flat elements) in tandem with 0th order functions as the temporal basis.  In this work, we develop the methodology necessary to extend to higher order surface descriptions as well as to enable its use with higher order temporal basis functions.  These higher order temporal basis functions are used in a Galerkin framework.  A number of results that demonstrate convergence, stability, and applicability are presented.

\end{abstract}

\section{Introduction}\label{Sec:Introduction}
Since the initial development of time domain integral equation (TDIE) solvers in the 1960s \cite{Friedman1962}, their use in electromagnetic simulations has only recently been on the uptick.  This renewed interest is thanks to the solution of the computational complexity bottleneck \cite{Yilmaz2004a,Ergin1999b} and increasingly sophisticated approaches to addressing instability.  While a plethora of stabilization schemes have been developed over the last two decades \cite{Rynne1990,Sadigh1993,Hu2001,Weile2004,Dodson1998,Sarkar2000}, the most promising schemes appear to be Lubich quadrature \cite{Lubich1988,Wang2011}, smooth and bandlimited temporal basis functions \cite{Weile2004}, quasi-exact integration \cite{Shanker2009,Shi2011}, and the separable expansion method \cite{Pray2012}.  Lubich quadrature relies on applying a Laplace transform to the appropriate TDIE and using a finite differencing scheme to map to the {\cal Z} transform domain.  This discretized equation is then solved by marching in time.  These methods have yielded excellent results, but are characterized by interactions with infinite temporal tails, leading to higher scaling in both computational cost and memory.  Methods to overcome this cost scaling are current topics of research.  Alternatively, smooth and bandlimited temporal basis functions can be effective in stabilizing TDIE solutions.  The smoothness of these functions allows for accurate evaluation of retarded potential integrals using conventional quadrature rules.  However, these functions are symmetric about their interpolation point and, therefore, produce marching systems that are noncausal, requiring extrapolation.  This necessitates the use of very small time steps to maintain stability.  Quasi-exact integration schemes transform surface integrals into volume integrals of the correlation function and the Green's function, which are then evaluated quasi-analytically.  The crux of this procedure is in determining the correlation function and its domain of support, which is difficult for higher order geometries.  This motivates the twin goals of this work, which are to develop a method that maintains temporal locality, while achieving higher order temporal accuracy and being extensible to higher order geometric descriptions.  As alluded to earlier, \cite{Pray2012} was a purely numerical approach to stabilizing TDIE solutions.  It is, therefore, easily extended to higher order descriptions of the geometry.  However, achieving high order accuracy in time is more challenging.

Historically, collocation in time has been the preferred method for time-marching schemes, although recent research has shown that Galerkin methods yield more accurate and stable methods \cite{Ha-Duong2003a,Beghein2012}.  To achieve higher order accuracy in time, in this work, we build on \cite{Beghein2012} and use a higher order temporal basis, albeit with a small variation-the support of these functions is restricted to one time interval.  This permits the use of Galerkin testing to develop a causal marching scheme.  The stability of the resulting time marching scheme is further improved by using the purely numerical approach developed in \cite{Pray2012}.  Additionally, separation of space-time allows for easy implementation of higher order discretization in space within this framework.

The paper is organized as follows:  Section \ref{Sec:Formulation} formulates the scattering problem, details its discretization, and presents modifications to impedance matrix elements using a separable representation; and Section \ref{Sec:Results} shows the interpolation accuracy of the new temporal basis and presents a number of scattering results, including convergence studies of farfield scattering.

\section{Formulation}\label{Sec:Formulation}
\subsection{Time domain EFIE, MFIE, and CFIE}
Consider a perfect electrically conducting (PEC) scatterer residing in free space.  Let $\partial\Omega$ represent the surface of the scatterer; the regions interior and exterior to the scatterer are denoted as $\Omega^{\intr}$ and $\Omega^{\ext}=\mathbb{R}^3\setminus \Omega^{\intr}$, respectively; the unit vector normal to $\partial\Omega$ is denoted as $\nhat$.  The scatterer is illuminated by an incident plane wave $\{\bE^{\inc}(\br,t),\bH^{\inc}(\br,t)\}$, which is assumed to be vanishingly small for time $t\leq0$ and effectively bandlimited to some frequency $\fmax$.  The incident field excites currents $\bJ(\br,t)$ on $\dOmega$, which in turn generate scattered fields $\{\bE^{\scat}(\br,t),\bH^{\scat}(\br,t)\},~(\br,t)\in\Omega^{\ext}\times[0,\infty)$ and the total fields in $\Omega^{\ext}$ are expressed as $\{\bE^{\tot}\spctime,\bH^{\tot}\spctime\}=\{\bE^{\inc}\spctime+\bE^{\scat}\spctime,\bH^{\inc}\spctime+\bH^{\scat}\spctime\}$.  The boundary conditions are given as
\begin{equation}
\begin{split}
&\left.\ncross\ncross\bE^{\tot}(\br,t)\right|_{\br\in\dOmega} = 0\\
&\left.\ncross\bH^{\tot}(\br,t)\right|_{\br\in\dOmega} = \bJ(\br,t)~.
\end{split}
\end{equation}
Expressing $\{\bE^{\scat}(\br,t),\bH^{\scat}(\br,t)\}$ in terms of $\bJ(\br,t)$ yields
\begin{equation}
\ncross \ncross\bE^{\inc}(\br,t) = {\cal L}\{\bJ(\br,t)\}
\label{TDEFIE}
\end{equation}
\begin{equation}
\ncross\bH^{\inc}(\br,t) = \bJ(\br,t) + {\cal K}\{\bJ(\br,t)\}
\label{TDMFIE}
\end{equation}
where
\begin{equation}
\begin{split}
{\cal L}\{\bJ(\br,t)\} &= \ncross\ncross\left\{\frac{\mu_0\dowt}{4\pi}\int_\dOmega d\br'\frac{\delta(t-R/c)}{R}\ast_t\bJ(\br',t)\right.\\-\frac{\nabla}{4\pi\eps_0}&\left.\int_\dOmega d\br'\frac{\delta(t-R/c)}{R}\ast_t\int_0^t dt'\Divp \bJ(\br',t')\right\}
\end{split}\label{Loperator}
\end{equation}
\begin{equation}
\begin{split}
{\cal K}\{\bJ(\br,t)\} = -\ncross\curl\frac{1}{4\pi}\int_\dOmega d\br'\frac{\delta(t-R/c)}{R}\ast_t\bJ(\br',t)
\end{split}\label{Koperator}
\end{equation}
where $\ast_t$ denotes the convolution operation in $t$, $R=\abs{\br-\br'}$, and the prime on $\nabla'$ denotes that the differentiation is with respect to $\br'$.  The integral in \eqref{Koperator} is computed in the principle value sense.  Both \eqref{TDEFIE} and \eqref{TDMFIE} suffer from the well known interior resonance problem for closed surfaces \cite{Shanker2000}.  This can be overcome by combining them as
\begin{equation}
\begin{split}
-\alpha/\eta_0\ncross &\ncross\bE^{\inc}(\br,t) + (1-\alpha)\ncross\bH^{\inc}(\br,t) =\\ -\alpha/\eta_0{\cal L}\{\bJ(\br,t)\}+&(1-\alpha)(\bJ\spctime+{\cal K}\{\bJ(\br,t)\})\doteq{\cal C}\{\bJ(\br,t)\}
\end{split}
\label{TDCFIE}
\end{equation}
where $\eta_0=\sqrt{\mu_0/\eps_0}$ is the intrinsic impedance of free space and $\alpha$ is a real number between 0 and 1.  The results presented in Section \ref{Sec:Results} will use a discrete version of either \eqref{TDEFIE} or \eqref{TDCFIE}.
\subsection{Space-Time Galerkin Discretization}\label{Subsec:STGalerk}
To solve either \eqref{TDEFIE}, \eqref{TDMFIE}, or \eqref{TDCFIE}, we begin by representing the current using a set of spatial and temporal basis functions as
\begin{equation}
\bJ(\br,t) = \sum_{n=1}^{\Ns} \bj_n(\br) \sum_{j=0}^{\Nt} \sum_{l=0}^\p I_n^{j,l}T_j^l(t)
\label{CURRENT_DEF}
\end{equation}
where $I_n^{j,l}$ are the unknown coefficients to be determined.  $\bj_n(\br)$ are RWG vector basis functions \cite{Rao1982}.  The temporal basis functions $T_j^l(t)$ are Lagrange polynomials, given as
\begin{equation}
\begin{split}
T_j^l(t) &= T^l(t-j\dt)\\
T^l(t) & = \begin{cases}{\displaystyle \ell_l(t)} \mbox{ } &t\in[-\dt,0]\\
{\displaystyle 0 } \mbox{ } &otherwise
\end{cases}\\
\ell_l(t) & = \prod_{\substack{i=0\\i\neq l}}^\p\frac{t-t_i}{t_l-t_i}\\
t_l & = (l/\p-1)\dt,~l=0,1,...,\p~.
\end{split}
\label{tbas}
\end{equation}
where $\dt$ is the time step.  Here we note that, as opposed to the standard definition of basis functions in, e.g. \cite{Shanker2009}, these functions have support over one time step, $\dt$.  This is essential to maintain causality in time marching when Galerkin testing is used in time.  On the contrary, basis functions that are typically used for MOT analyses \cite{Shanker2009,Weile2004} cannot be used within a Galerkin framework, while maintaining causality, as shown in the appendix.  We note here that the temporal basis functions in \eqref{tbas} are discontinous between time segments, i.e. at $t=j\dt$.  While this does not present any difficulties in discretizing the TD-EFIE and TD-MFIE, this testing/basis set cannot be used in discretizing the time differentiated forms of TD-EFIE and TD-MFIE because the correlation function resulting from Galerkin testing is continuous, but not smooth.  To discretize the time differentiated EFIE and/or MFIE, the functions outlined in \cite{Beghein2012} should be used.  This claim is justified in the appendix.

Substituting \eqref{CURRENT_DEF} into \eqref{TDCFIE}, and employing Galerkin testing in both space and time yields

\begin{equation}
	 \sum_{j=0}^{i} \underline{{\underline{Z_{i-j}}}} ~\underline {I_{j}}  = \underline {V_i}
	\label{tmat}
\end{equation}
where 
\bml
\begin{equation}
	\underline{I_j} = \left[I_1^{j,0} ~I_1^{j,1} ...~I_{\Ns}^{j,\p}\right]^T
	\underline{V_i} = \left[v_1^{i,0} ~v_1^{i,1} ...~v_{\Ns}^{i,\p}\right]^T
\end{equation}
\begin{equation}
	\begin{split}
	v_m^{i,k}=&-\alpha/\eta_0\inprod{\bj_m(\br)T_i^k(t)}{\ncross \ncross \bE^{\inc}(\br,t)}\\
		&+ (1-\alpha)\inprod{\bj_m(\br)T_i^k(t)}{\ncross \bH^{\inc}(\br,t)}
	\end{split}
\end{equation}
\begin{equation}
	\begin{split}
	Z_{i-j,kl}^{mn}&=\langle\bj_m(\br)T_i^k(t),{\cal C}\{\bj_n(\br)T_{j-i}^l(t)\}\rangle
	\end{split}
\label{testedeq2}
\end{equation}
\begin{equation}
\underline{{\underline{Z_{i-j}}}}=
\left(
\begin{tabular}{cccccc}
$Z_{i-j,00}^{11}$ & $Z_{i-j,01}^{11}$ & ... & $Z_{i-j,0\p}^{11}$ & ... &  $Z_{i-j,0\p}^{1\Ns}$\\
$Z_{i-j,10}^{11}$ &  &  &  &  &  \\
. &  &  & . &  &  \\
. &  &  & . &  & \\
. &  &  & . &  &  \\
$Z_{i-j,\p0}^{11}$ &  & ... & $Z_{i-j,\p\p}^{11}$  &  &  \\
. &  &  &  &  & .\\
. &  &  &  &  & .\\
. &  &  &  &  & .\\
$Z_{i-j,\p0}^{\Ns1}$ &  &  &  & ... & $Z_{i-j,\p\p}^{\Ns\Ns}$
\end{tabular}
\right)
\label{Zmat}
\end{equation}
\eml
The dimensions of $\underline{{\underline{Z_{i-j}}}}$ are $\Ns(\p+1)\times \Ns(\p+1)$ as opposed to $\Ns\times \Ns$ as in the case of a traditional temporal basis set.

\subsection{Evaluation of Inner Products}\label{Subsec:SepExp}
\begin{figure}[h!]
	\begin{center}
       \begin{tikzpicture}[decoration=brace]
	       \draw (-1.75,-2) -- (-1,1) node[above] {\small \bf Observation };
	       \draw (-1,1) -- (0.25,-0.5);
	       \draw (0.25,-0.5) -- (-1.75,-2);
	       \draw (2.7,1.45) -- (5,-.8);
	       \draw (5,-.8) -- (3.15,-.85) node[below,midway] {\small \bf Source};
	       \draw (3.15,-.85) -- (2.7,1.45);
	       \draw[->] (-.7,-.5) -- (3.84254,-.0457413) ;
	       \draw[<->] (1.77799cm,-0.49cm) -- (2.952cm,-.37cm) node[below,midway] {\small $c\Delta t$};
	       \draw[decorate] (-.8,-.4) -- (2.8575,-.05) node[above,midway] {\small $\zeta$} ;
	       \draw[decorate]  (2.963,-0.0299) -- (3.843,.0543) node[above,midway] {\small $R-\zeta$} ;
		\draw (.48933cm,-.80292cm) arc (-14.289:25.711:1.2273cm);
		\draw (1.6787cm,-1.1058cm) arc (-14.289:25.711:2.4546cm);
		\draw (2.8680cm,-1.4088cm) arc (-14.289:25.711:3.68191cm);
		\draw (4.8721cm,-1.9192cm) arc (-14.289:25.711:5.75cm);
	        \draw[<->] (2.84436cm,0.49699cm) -- (4.8352cm,1.0570cm) node[above,midway] {\small $\beta c\Delta t$};
		\draw  (-.8cm,-.7cm) node[left] {\small \bf ${\bf r}$}; 
		\draw  (3.94cm,-.02cm) node[below] {\small \bf ${\bf r}'$}; 
	       \pgfplothandlermark{\pgfuseplotmark{x}}
		\pgfplotstreamstart
		\pgfplotstreampoint{\pgfpoint{-.8cm}{-0.5cm}} node[below] 
		\pgfplotstreampoint{\pgfpoint{3.92806cm}{-.0376cm}}
		\pgfplotstreamend
		\pgfusepath{stroke}
	 \end{tikzpicture}
	 \caption{\label{tricuts} Separable expansion parameters}
	\end{center}
\end{figure}
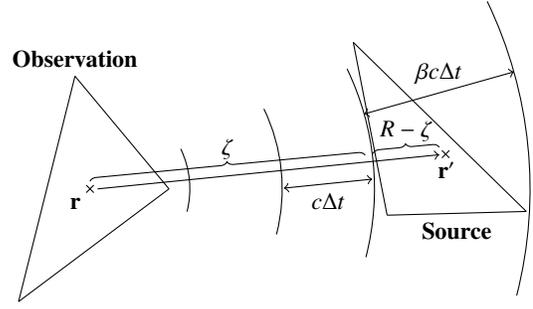

The integrand in \eqref{testedeq2} is piecewise continuous and application of standard quadrature rules to \eqref{Loperator} and \eqref{Koperator} does not lead to accurate evaluation of the integrals.  One needs to explicitly account for the discontinuities in the integrand.  This can be achieved via quasi-exact integration \cite{Shanker2009,Shi2011}, i.e. by identifying regions where the integrand is smooth and analytically integrating over each subregion.  However, as discussed in Section \ref{Sec:Introduction}, this procedure is restricted to first order spatial discretizations.  This restriction leads us to adopt the approach of \cite{Pray2012}, which is summarized here for the sake of completeness.  To begin, we define the vector potential as
\begin{equation}
{\bf A}\{\bJ\spctime\}=\int_\dOmega d\br'\frac{\bJ(\br',\tau)}{4\pi R}
\end{equation}
where $\tau=t-R/c$.  Employing space-time Galerkin testing to the contribution of the vector potential due to a source $\bj_n(\br)T_j^l(t)$ yields
\begin{equation}
	\begin{split}
	&\langle\bj_m(\br)T_i^k(t),\dow_t{\bf A}\{\bj_n(\br)T_j^l(t)\}\rangle\\
	&=\int_\dOmegam d\br\bj_m(\br)\cdot\int_\dOmegan d\br'\bj_n(\br')\int_{(i-1)\dt}^{i\dt}dtT_i^k(t) \frac{\dowt T^l_j(\tau)} {4 \pi R}\\
	&\doteq\int_\dOmegam d\br\bj_m(\br)\cdot\int_\dOmegan d\br'\bj_n(\br')\psi_{ij,kl}({\bf r,\bf r'})
	\end{split}
\label{testedLop}
\end{equation}
where $\dOmegan$ is the support of $\bj_n(\br)$.  Next, we make the approximation
\begin{align} 
\label{eq:SepExp}
 \psi_{ij,kl}({\bf r},{\bf r}')  &=   \int_{(i-1)\dt}^{i\dt}dtT_i^k(t)\delta \left( t - \frac{\zeta}{c}\right) \nonumber\\&~~~~~\ast_t \frac{\delta \left ( t - \frac{R-\zeta}{c} \right )}{4 \pi R}  \ast_t \dowt T^l_j(t)\nonumber\\ 
 ~~~~&= \frac{1}{4 \pi R} \sum_{q=0}^\infty a_q P_q \left(\xi\right) \tilde{T}_{ij,kl}^q\\ 
 &\approx \frac{1}{4 \pi R}\sum_{q=0}^{N_h} a_q P_q \left(\xi\right) \tilde{T}_{ij,kl}^q\nonumber\\
 \text{where}~\xi = k_1 &(R-\zeta)/c + k_2 ,~a_q = k_1\frac{2 q + 1}{2},\nonumber\\\tilde{T}_{ij,kl}^q&=\int_{(i-1)\dt}^{i\dt}dtT_i^k(t)\delta \left (t - \frac{\zeta}{c}\right ) \ast_t \nonumber\\&~~~~~~~~~~P_q(k_1 t + k_2 ) {\cal P}_{0,\beta}(t) \ast_t \dowt T_j^l(t)~.\nonumber
\end{align}
Here, $P_q(k_1t+k_2)$ is a Legendre polynomial of order $q$ with support $t\in[0,\beta\dt]$, $k_1$ and $k_2$ are real numbers providing the mapping $[0,\beta\dt]\rightarrow[-1,1]$, $\zeta$ is the maximum multiple of $c\dt$ between $\br$ and the source triangle (see figure \ref{tricuts}), and ${\cal P}_{0,\beta}(t)$ is a window function defined as
\begin{equation}
{\cal P}_{0,\beta}(t)=\begin{cases} 1 & t\in[0,\beta\dt]\\
0 & otherwise~.\end{cases}
\end{equation}
The expansion in \eqref{eq:SepExp} enables accurate evaluation of the spatial integrals in \eqref{testedeq2}.  It can be shown that the expansion \eqref{eq:SepExp} is uniformly convergent.  For a given number of harmonics, $\Nh$, an expression for the error bound is given in the appendix (the same expansion can be used in computing the tested gradient of the scalar potential).

Next, we present the modifications to the matrix elements in \eqref{testedeq2} using the approximation \eqref{eq:SepExp}.  Continuing from \eqref{testedLop}, the contribution from the time differentiated vector potential is given by 
\bml
\begin{equation}
\label{eq:ApotSep}
\left < {\bj}_m ({\bf r})T_i^k(t),\dowt{\bf A}\left\{ {\bj}_n({\bf r}) T_j^l(t)\right \}\right >  \approx\frac{\mu_0}{4 \pi} \sum_{q=0}^{N_h} a_q \alphaq  \tilde{T}_{ij,kl}^q
\end{equation}
where
\begin{equation}
	\displaystyle \alphaq  = \int_{\dOmegam} d{\bf r} {\bj}_m ({\bf r})  \cdot\int_{\dOmegan} d{\bf r}' \frac{{\bj}_n ({\bf r}')P_q \left(\xi\right)}{R} 
	\label{eq:ASpat}
\end{equation}
\eml

Next, we define the scalar potential as
\begin{equation}
{\Phi}\{\bJ\spctime\}=-\int_\dOmega d\br'\int_{-\infty}^t dt'\frac{\Divp\bJ(\br',t'-R/c)}{4\pi R}~.
\end{equation}
Using the expansion \eqref{eq:SepExp} the contribution of this quantity to \eqref{testedeq2} is
\bml
\begin{equation}
\label{eq:PpotSep}
\left < {\bj}_m ({\bf r})T_i^k(t),\grad{\Phi}\left\{ {\bj}_n({\bf r}) T_j^l(t) \right \}\right >  \approx\frac{1}{4 \pi\eps_0} \sum_{q=0}^{N_h} a_q \phiq  \hat{T}_{ij,kl}^q
\end{equation}
where
\begin{equation}
	\displaystyle \phiq  = \int_{\Omega_m} d{\bf r} \Div{\bj}_m ({\bf r})  \int_{\Omega_n} d{\bf r}' \frac{\Divp{\bj}_n ({\bf r}')P_q \left(\xi\right)}{R} 
	\label{eq:PSpat}
\end{equation}
\begin{equation}
	\begin{split}
	\displaystyle \hat{T}_{ij,kl}^q = \int_{(i-1)\dt}^{i\dt}&dtT_i^k(t)\delta \left (t - \frac{\zeta}{c}\right ) \ast_t \\&P_q(k_1 t + k_2 ) {\cal P}_{0,\beta}(t) \ast_t \int_{-\infty}^t dt' T_j^l (t')~.
	\end{split}
\end{equation}
\eml
Lastly, the tested operator ${\cal K}\{\bJ(\br,t)\}$ becomes
\bml
\begin{equation}
\label{eq:HSep}
 \left < {\bj}_m ({\bf r})T_i^k(t),{\cal K}\left\{ {\bj}_n({\bf r}) T_j^l(t) \right \}\right >  \approx\frac{1}{4 \pi} \sum_{q=0}^{N_h} a_q \kappaq  \bar{T}_{ij,kl}^q
\end{equation}
where
\begin{equation}
	\begin{split}
	\displaystyle \kappaq  = \int_{\Omega_m} d{\bf r} &{\bj}_m ({\bf r}) \cdot \ncross\int_{\Omega_n} d{\bf r}' {\bj}_n ({\bf r}')\\&\left(\frac{k_1P_q' \left(\xi\right)}{cR}
	-\frac{P_q\left(\xi\right)}{R^2}\right)\times\hat{R} 
	\end{split}
	\label{eq:HsSpat}
\end{equation}
\begin{equation}
	\begin{split}
	\displaystyle \bar{T}_{ij,kl}^q = \int_{(i-1)\dt}^{i\dt}&dtT_i^k(t)\delta \left (t - \frac{\zeta}{c}\right ) \ast_t \\&P_q(k_1 t + k_2 ) {\cal P}_{0,\beta}(t) \ast_t T_j^l (t)~.
	\end{split}
\end{equation}
\eml
The prime on $P_q'(\cdot)$ denotes the 1st derivative of $P_q(\cdot)$ with respect to its argument.  Returning to \eqref{eq:PpotSep}, it can be shown that, for $(i-j)>\zetan+\beta + 1$,
\begin{equation}
\begin{split}
\hat{T}_{ij,kl}^q&=\frac{1}{a_0}A_{kl}\delta_{q0}\\
 A_{kl} &= \left(\int_{(i-1)\dt}^{i\dt}dtT_i^k(t)\right)\left(\int_{(j-1)\dt}^{j\dt} dt T_j^l (t)\right)
\end{split}\label{eq:PhiSplit}
\end{equation}
where $\zetan=\zeta/(c\dt)$ and $\delta_{mn}$ is the Kronecker delta.  Substituting \eqref{eq:PhiSplit} into \eqref{testedeq2} yields the modified marching system
\bml
\begin{equation}
	 \sum_{j=0}^{i} \underline{{\underline{Z_{i-j}}}} ~\underline {I_{j}} + \sum_{j=0}^{i} \underline{{\underline{Z^\phi_{i-j}}}} ~\underline {I_{j}}  = \underline {V_i}
	\label{tmat2}
\end{equation}
\begin{equation}
	\left[\underline{{\underline{Z_{i-j}^{^\phi,mn}}}}\right]_{k,l}=\begin{cases}
	\frac{1}{4 \pi\eps_0} \phizero A_{kl} & i-j >\beta+1+\zetan\\
	0 & i-j \le\beta+1+\zetan\end{cases}~.
\label{testedphieq2}
\end{equation}
Therefore,
\begin{equation}
\begin{split}
	 \sum_{j=0}^{i} \left[\underline{{\underline{Z^{\phi,mn}_{i-j}}}}\right]_{k,l} ~I_n^{j,l} &= \sum_{j=0}^{i-\beta-2} \left(\frac{\phizero A_{kl}}{4 \pi\eps_0} \right)I_n^{j,l}\\
	 &=\frac{\phizero A_{kl}}{4 \pi\eps_0} C_n^{i-\beta-2,l}
\end{split}
\end{equation}
where
\begin{equation}
C_n^{i,l} = \sum_{j=0}^{i}I_n^{j,l}~.
\label{eq:recursive_charge}
\end{equation}
\eml
\eqref{eq:recursive_charge} is used while marching in time to recursively compute the charge terms $C_n^{i,l}$ while retaining the ${\cal O}(\Nt\Ns^2)$ scaling of the solver.

The integrals in \eqref{eq:ASpat}, \eqref{eq:PSpat}, and \eqref{eq:HsSpat} are of the form
\begin{equation}
\int_{\Omega_n}d\br'\frac{\bj(\br')P_q(\xi)}{R^\alpha}
\end{equation}
which can be evaluated to arbitrary precision using an appropriate order of integration rule.  Obviously, if $q$ is too high then the order of integration will, of necessity, be impractically high.  The order of integration is determined by $N_h$, i.e. the highest order harmonic required for an interaction pair.  While theoretical bounds on $N_h$ can be obtained for a given error (see Section \ref{subsec:Nhconv}), this is not a tight estimate.  In practice $N_h$ is significantly lower.  This is elucidated further in the next Section.

\section{Results}\label{Sec:Results}
\subsection{Interpolation}

\begin{figure}[!h]
     \centering
     	\includegraphics[width=\linewidth]{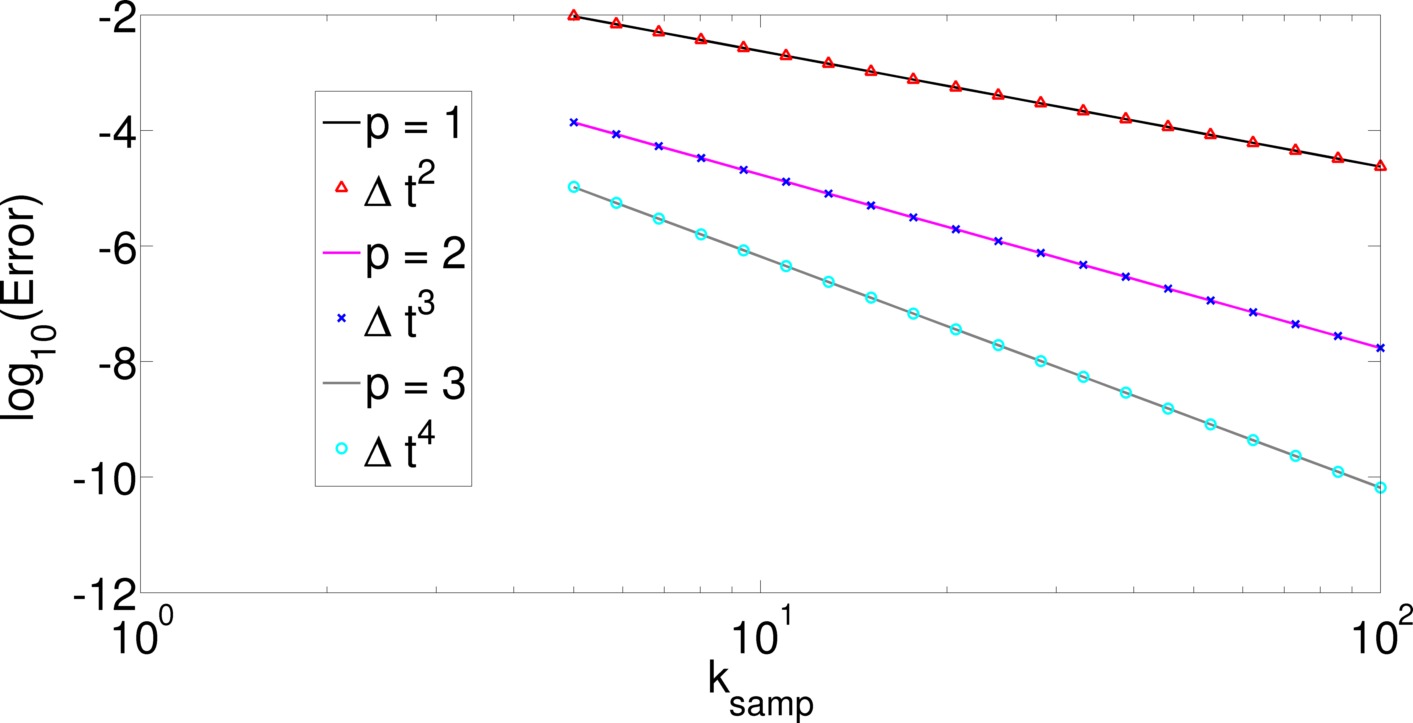}
     	\caption{Scaling of interpolation error for various $\p$ and sampling frequencies}
	\label{interp_error}
\end{figure}

To test the accuracy of these basis functions, they are used to interpolate and shift an approximately bandlimited modulated gaussian pulse of the form 
\begin{equation}
f(t) = \text{cos}(\omega_0 t)e^\frac{-(t-tp)^2}{2\sigma^2}
\end{equation}
where $\sigma=3/(2\pi \fmax)$, $t_p=6\sigma$, and $\fmax$ denotes the frequency at which the power is approximately 160 dB below the peak value at $f_0=\omega_0/2\pi$.  Figure \ref{interp_error} shows the order of interpolation accuracy for various $\p$.  The time step is chosen as $\dt=1/\left(2\ksamp\fmax\right)$, where $\ksamp>0$ is a real number. The error is defined as $\text{Error}=\norm{f_{approximate}-f_{analytic}}/\norm{f_{analytic}}$ where,
\begin{equation}
\begin{split}
f_{analytic} = &[f(\dt-\Delta)~f(2\dt-\Delta) ~... ~f(\Nt\dt-\Delta)] \\
f_{approximate} = &[\tilde{f}(\dt-\Delta) ~\tilde{f}(2\dt-\Delta) ~... ~\tilde{f}(\Nt\dt-\Delta)] \\
 &\tilde{f}(t) = \sum_{j=1}^{\Nt}\sum_{l=0}^pI_j^lT_j^l(t)~.
\end{split}
\end{equation}
The coefficients $I_j^l$ are computed by approximating an unshifted Gaussian and applying Galerkin testing.  $\Delta/\dt$ is defined as some shift between 0 and 1, which in this experiment is set to 0.906.  The center frequency is set to $f_0=200$ MHz and $\fmax=300$ MHz.
As expected, the error can be seen in figure \ref{interp_error} to scale as ${\cal O}(\dt^{\p+1})$ for the different values of $\ksamp$.

In order to estimate the lower bound of $\Nh$ we repeat this interpolation experiment for various basis function orders and time step sizes.  For each setup, we compute the number of harmonics needed to reproduce the interpolation accuracy of the same basis function order/time step size achieved without the approximation \eqref{eq:SepExp}.  The value of $\beta$ was chosen to mimic a typical simulation on a tessellation consisting of elements approximately $\lambda_\text{min}/10$ in length, i.e.
\begin{equation}
\begin{split}
\beta &= \ceil{\frac{\lambda_\text{min}/10}{c\dt}}\\
&=\ceil{\ksamp/5}
\end{split}
\end{equation}
where $\lambda_\text{min}=c/\fmax$.
 \begin{table}
   \small
   \centering
   \begin{tabular}{|c|c|c|c|c|} \hline
   \backslashbox[5pt]{$\p$}{$\ksamp$} & $5$ & $10$ & $20$ &  $40$\\ \hline
     1   &   1/1/1    &    1/1/2   &   2/1/2   &  3/1/3 \\ \hline 
     2   &   2/1/1    &    4/2/2   &   4/3/3   &  4/3/3 \\ \hline 
     3   &   4/3/4    &    4/4/4   &   6/4/5   &  7/5/5 \\ \hline 
\end{tabular}
\caption{Lower bound of $\Nh$ needed to reproduce interpolation accuracies (undifferentiated/1st derivative/integral)}
\label{interp_Nh}
\end{table}
Table \ref{interp_Nh} tabulates the minimum value of $\Nh$ required to produce the error levels provided in figure \ref{interp_error}.
\subsection{Scattering Results}
The remainder of this Section presents scattering results from a variety of scatterers.  For each simulation, a PEC scatterer is illuminated by a plane wave of the form
\begin{equation}
	{\bf E}^{\inc}({\bf r},t)={\hat u}\cos(2\pi f_0t)e^{-(t-{\bf r}\cdot{\hat k} /c-t_p)^2/2\sigma^2 }
	\label{einc}
\end{equation}
where ${\hat u}$ is the electric polarization and ${\hat k}$ denotes the direction of propagation.  $\sigma$, $t_p$, and $f_0$ are defined in Section \ref{Subsec:STGalerk}.  For all simulations, except where noted to the contrary, $\dt=1/(20\fmax)$ ($\ksamp=10$) and $\p=2$.  RCS comparisons are made at $f_0$, $f_0+\df$, and $f_0-\df$, where $f_0+\df<\fmax$.  The error against a frequency domain solver (or analytical result where available),
\begin{equation}
\text{Error}=\frac{\norm{\zeta_{\tdie}-\zeta_{\fdie}}}{\norm{\zeta_{\fdie}}}
\label{eq:rcs_error}
\end{equation}
is given at these three frequencies, while the plot is shown for only $f_0$ and $f_0+\df$ in order to maintain a reasonable range.  The $\ell^2$ norm is used.  Here $\RCS_{\tdie}=10~\text{log}_{10}(\zeta_{\tdie})$ and $\RCS_{\fdie}=10~\text{log}_{10}(\zeta_{\fdie})$ denote the radar cross sections obtained at a discrete set of angles using a TDIE and FDIE solver, respectively.

\begin{figure}[h!]
     \centering
     \includegraphics[width=\linewidth]{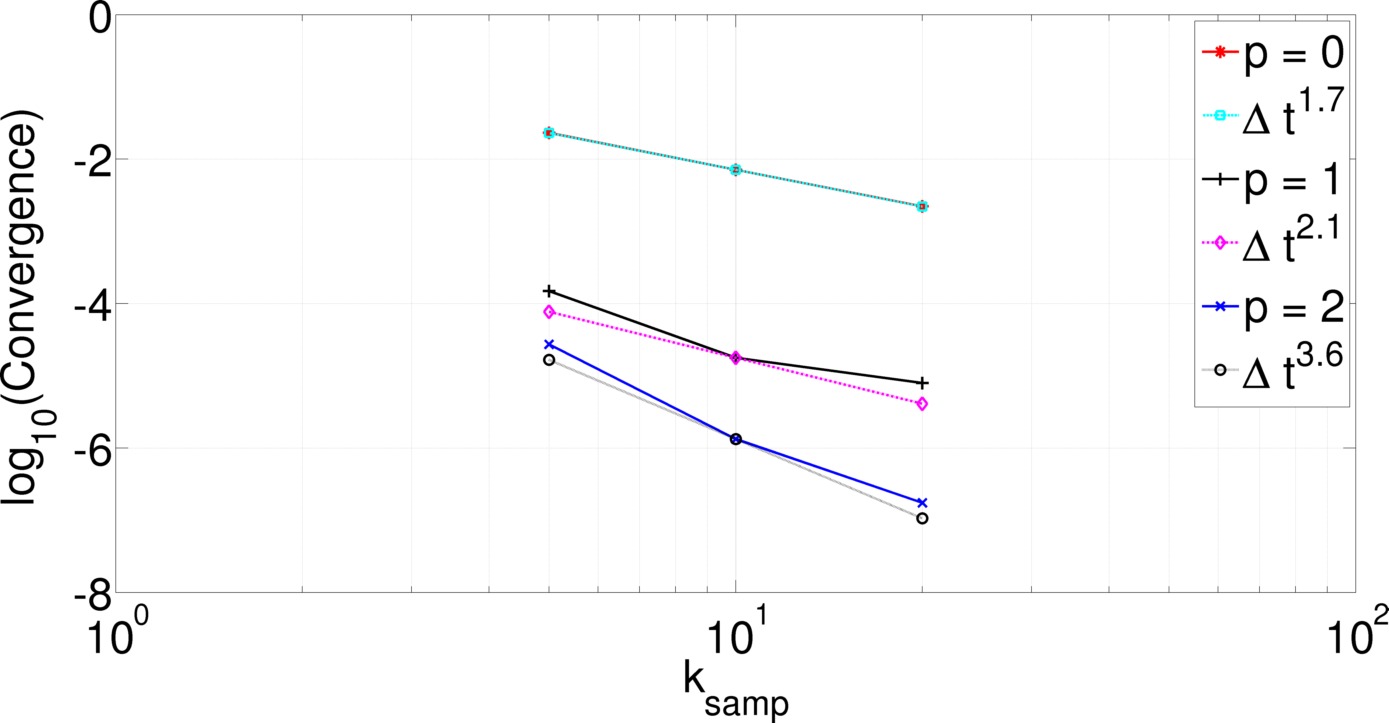}
     \caption{RCS convergence for plate}
\label{Etheta_conv_plate}
\end{figure}

For our first result, we examine convergence in temporal basis function order and sampling frequency.  A plate of dimensions 1m $\times$ 1m, discretized using 200 triangular elements, is illuminated by a plane wave of parameters $f_0=150$ MHz, $\fmax=225$ MHz, ${\hat k}={\hat z}$, and ${\hat u}={\hat x}$.  The time domain simulation was performed for three values of $\p$ and four values of $\ksamp$.  For this result, we do not compare against frequency domain results, but rather look at convergence in farfield scattering.  Here, convergence is defined for given values of $\theta$ and $\phi$ as 
\begin{equation}
\text{Convergence}_{k+1} = \frac{\abs{\Etheta_{k+1}(\theta,\phi)-\Etheta_{k}(\theta,\phi)}}{\abs{\Etheta_{k+1}(\theta,\phi)}}
\label{eq:Etheta_convergence}
\end{equation}
where ${\bf E}^{\scat}(\theta,\phi)\approx\phihat\Ephi_k(\theta,\phi)+\thetahat\Etheta_k(\theta,\phi)$ is the farfield approximation to the scattered field for a given sampling frequency, indexed by $k$.  The result is shown in figure \ref{Etheta_conv_plate} for an angle of $\theta=-130\degree$ and $\phi=0\degree$.  The rates of convergence were seen to be nonuniform across observation angles $\theta$ in the $x-z$ plane, and were found to lie roughly between ${\cal O}(\dt^{\p+1})$ and ${\cal O}(\dt^{\p+2})$.  

\begin{figure}[h!]
     \centering
     \includegraphics[width=\linewidth]{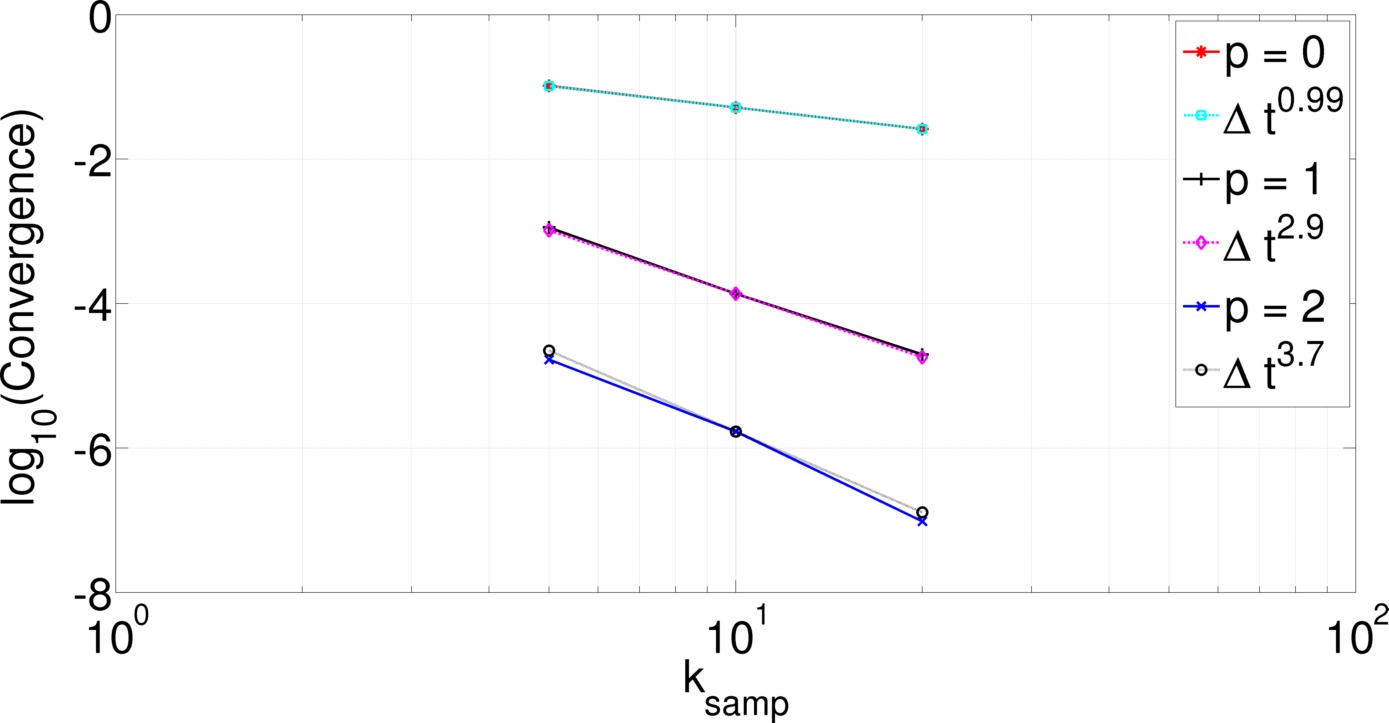}
     \caption{RCS convergence for sphere}
\label{Etheta_conv_sphere}
\end{figure}

For our second result, we perform the same convergence study, but for scattering from a sphere of radius $1$ m, discretized with $576$ unknowns.  The sphere is illuminated by a plane wave with $f_0=60$ MHz, $\fmax=90$ MHz, $\hat{u}=\hat{x}$, and $\hat{k}=\hat{z}$.  The TD-CFIE is used with $\alpha=0.5$ and the farfield is computed for $\theta=-180\degree$ and $\phi=0\degree$.  Figure \ref{Etheta_conv_sphere} shows the convergence in the farfield for the various simulation setups. 
Similarly to the previous result, the rates of convergence for various values of $\theta$ in the $x-z$ plane were found to lie between ${\cal O}(\dt^{\p+1})$ and ${\cal O}(\dt^{\p+2})$.

Our next scatter is a thin PEC box, discretized using $1,146$ unknowns, with dimensions $100\times50\times10$ m.  The scatterer is excited by a plane wave with parameters $f_0=1.4$ MHz, $\fmax=2.7$ MHz, $\hat{u}=\hat{y}$, and $\hat{x}=\hat{k}$.  This is a difficult scattering problem to stabilize due to the relatively small dimensions in the $z$-direction, particularly so using the TD-EFIE. 
\begin{figure}[!h]
     \centering
     \includegraphics[width=\linewidth]{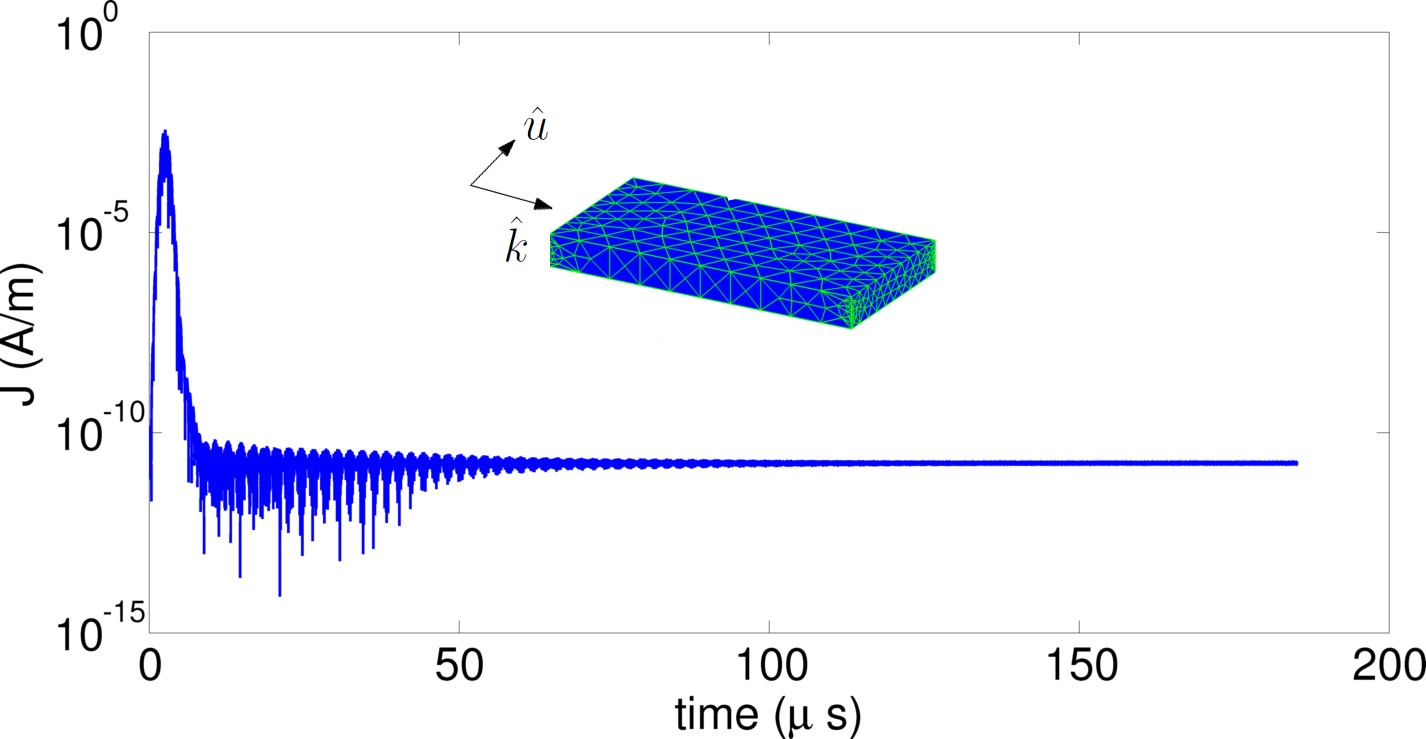}
     \caption{Current coefficient on box}
	\label{thbox_cur}
\end{figure}
 The current is observed for $10,000$ time steps (57 transits across geometry) and can be seen in figure \ref{thbox_cur} to remain stable throughout the duration of the simulation.
\begin{figure}[!h]
     \centering
     \includegraphics[width=\linewidth]{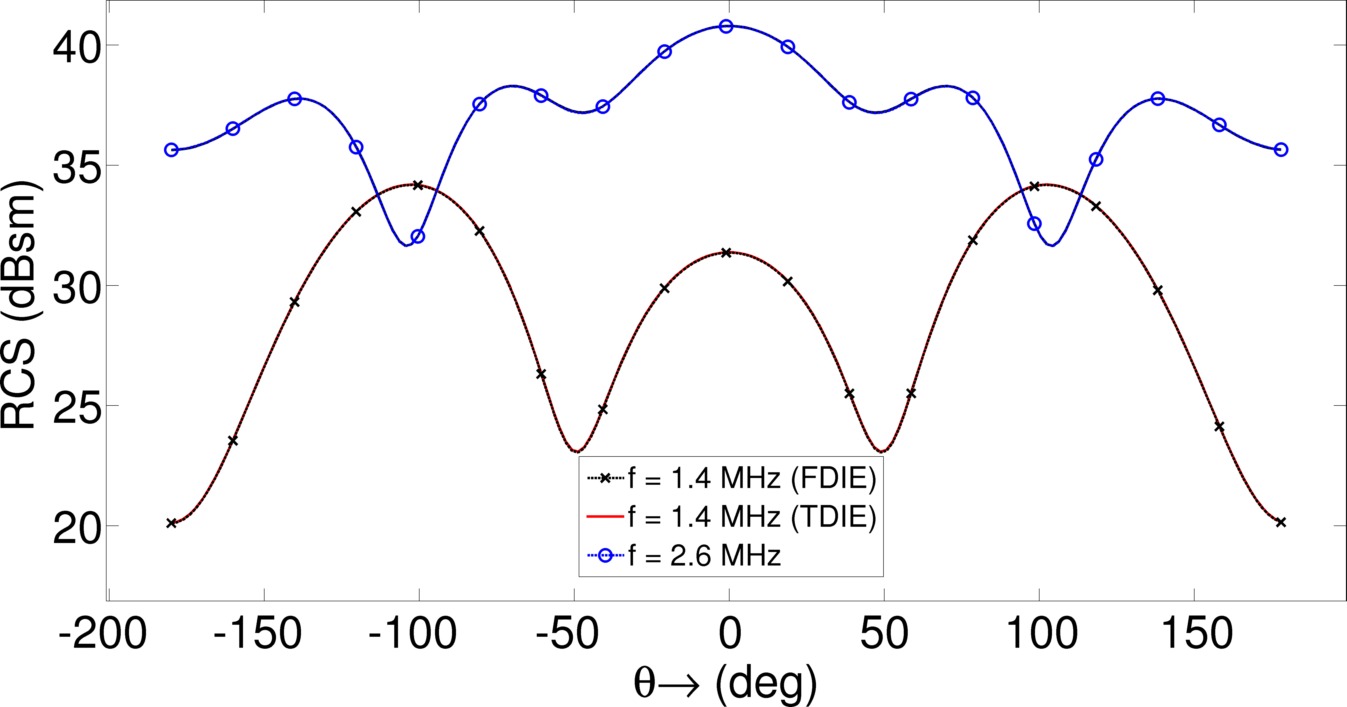}
     \caption{RCS of box}
	\label{thbox_rcs}
\end{figure}
To validate the accuracy of the solution, the RCS is computed in the $x-y$ plane and compared with a validated frequency domain EFIE solver.  The radar cross section of the box is obtained from the two solvers at 3 different frequencies.  Figure \ref{thbox_rcs} shows the RCS values at the two higher frequencies and close agreement is seen.  The error between the two solutions was found using \eqref{eq:rcs_error} to be 0.40\%, 0.36\%, and 0.19\% at the 0.2, 1.4, and 2.6 MHz, respectively.

Our next result is scattering from a cone-sphere of $7,965$ unknowns discretized with 2nd order elements with its axis along the $z$-direction.  The spherical portion of the scatterer has radius $1$ m, while the height of the cone is $4$ m.  The excitation has $f_0=80$ MHz, $\fmax=150$ MHz, $\hat{u}=\hat{y}$, and $\hat{k}=x$.  The current for $4,000$ time steps (80 transits across geometry) is computed using the TD-CFIE with $\alpha=0.5$.
\begin{figure}[!h]
     \centering
     \includegraphics[width=\linewidth]{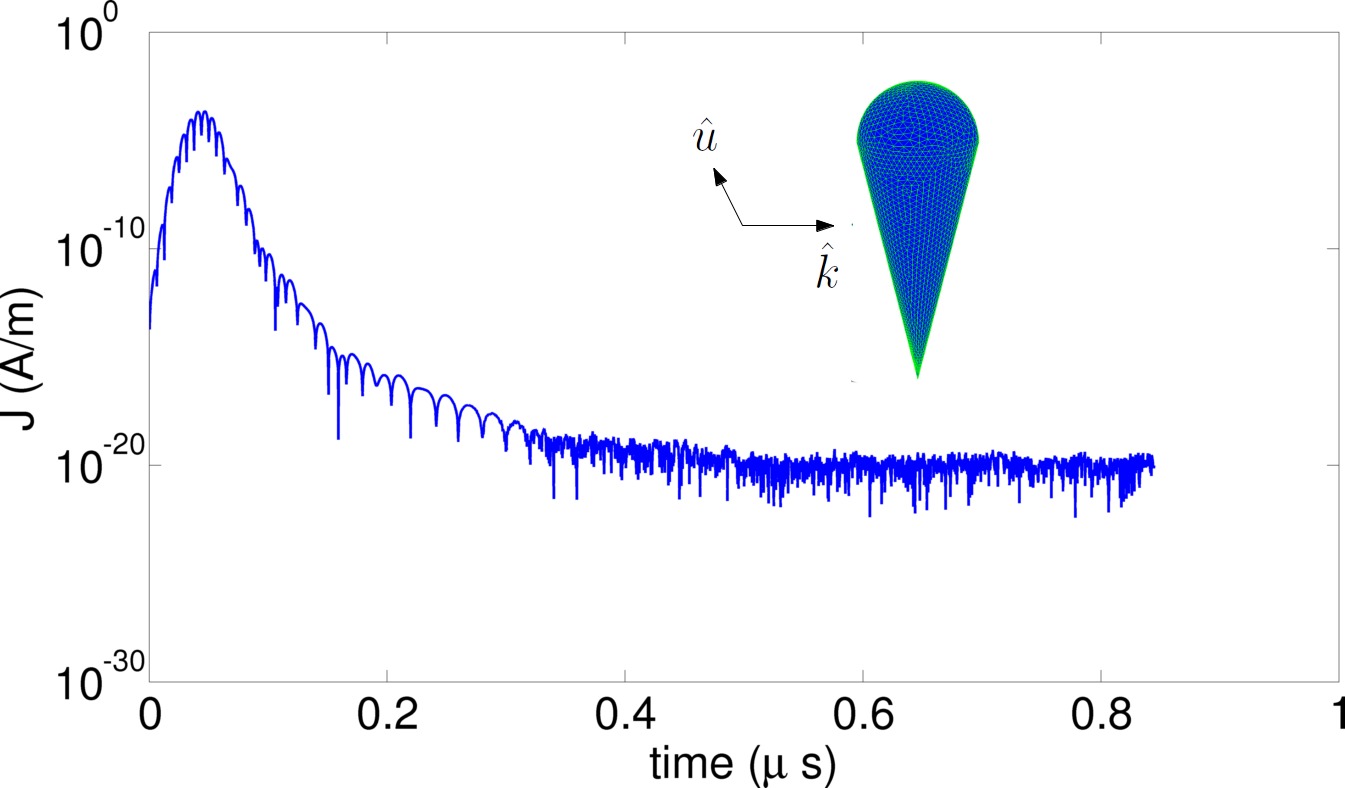}
     \caption{Current coefficient on cone-sphere}
	\label{cnsphr_cur}
\end{figure}
\begin{figure}[!h]
     \centering
     \includegraphics[width=\linewidth]{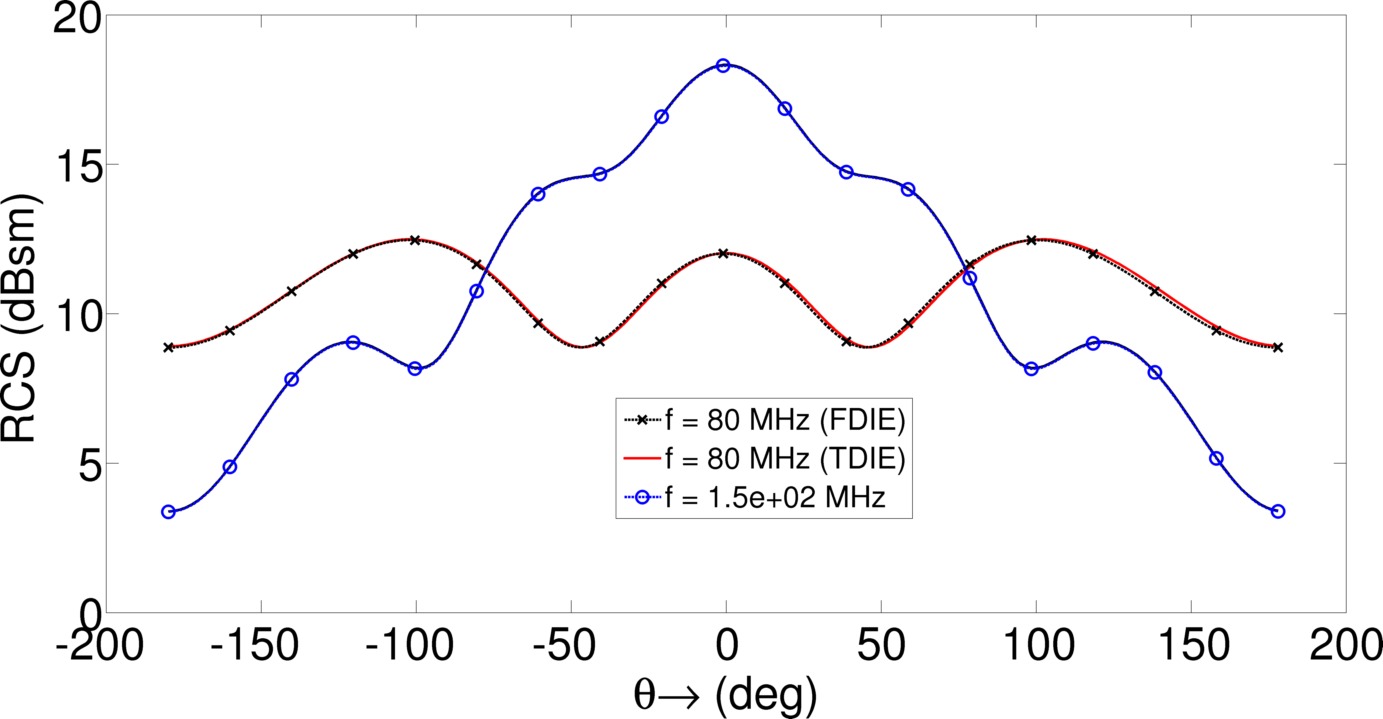}
     \caption{RCS of cone-sphere}
	\label{cnsphr_rcs}
\end{figure}
This is a challenging geometry due to the sharp tip of the scatterer, yet as is seen in figure \ref{cnsphr_cur}, the current remains stable throughout the simulation.  The solution is again validated in figure \ref{cnsphr_rcs} via RCS comparison with an FD-CFIE solver.  The RCS is computed in the $x-y$ plane and, again, close agreement betewen the time and frequency domain solutions is seen, with the errors given as 0.29\%, 1.7\%, and 0.49\% at 11, 80, and 149 MHz, respectively.

Our next result is a thin almond shaped scatterer of $1,668$ unknowns, discretized with 2nd order elements.  The geometry fits within a box of $5\times4.33\times0.865$ m.  The incident wave has parameters $f_0=20$ MHz, $\fmax=35$ MHz, $\hat{u}=\hat{x}$, and $\hat{k}=\hat{z}$.  The TD-CFIE with $\alpha=0.5$ is used.
\begin{figure}[!h]
     \centering
     \includegraphics[width=\linewidth]{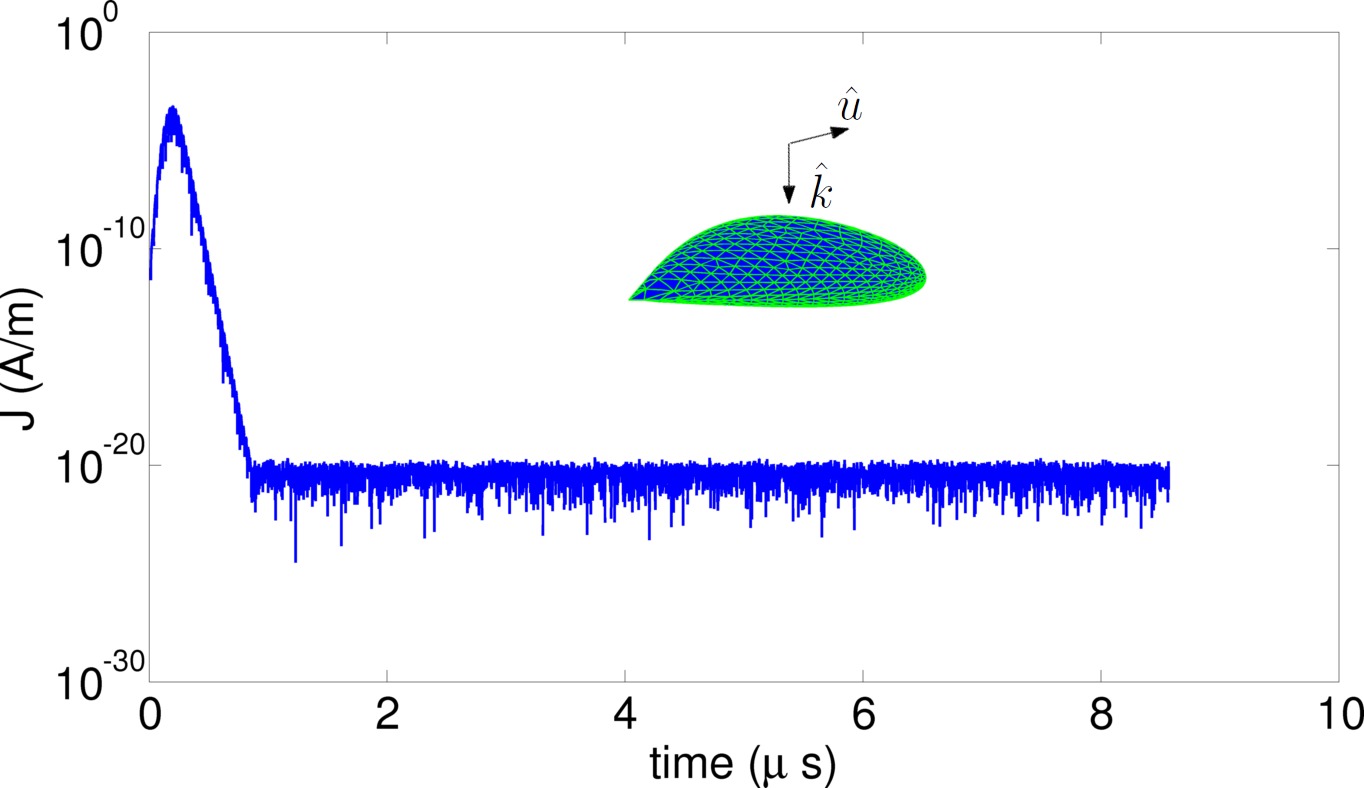}
     \caption{Current coefficient on almond shaped scatterer}
	\label{alm_cur}
\end{figure}
\begin{figure}[!h]
     \centering
     \includegraphics[width=\linewidth]{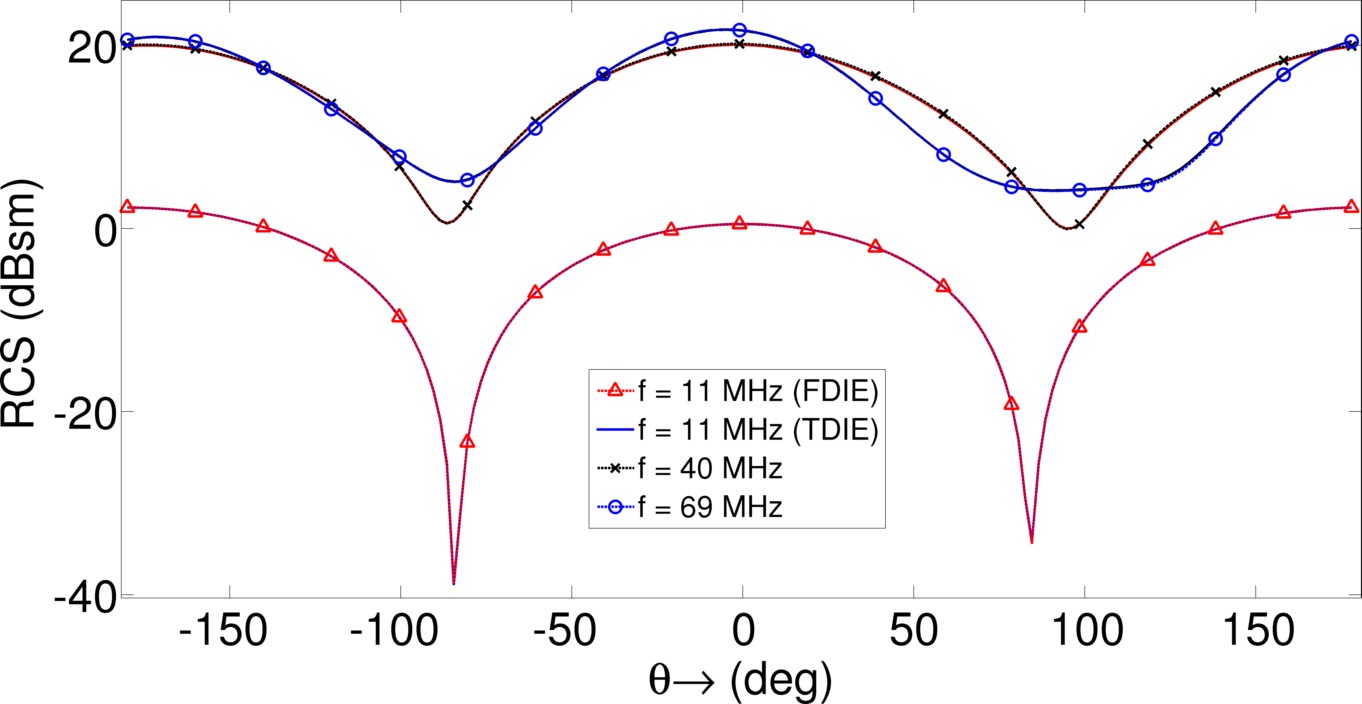}
     \caption{RCS of almond shaped scatterer}
	\label{alm_rcs}
\end{figure}
Figure \ref{alm_cur} shows that the current remains bounded throughout the simulation (80 transits across geometry or $6,000$ time steps).  This geometry is challenging given the sharp tip and the fact that it is very thin in the $z$-direction.  Validation is presented in figure \ref{alm_rcs} via comparison of RCS in the $x-z$ plane with an FD-CFIE solver.  The error is found to be 0.038\%, 2.1\%, and 0.59\% at 11, 40, and 69 MHz, respectively.

Our final result is solution to the EFIE for scattering from a VFY218.  This structure fits in a box of size $15\times9\times4$ m and is discretized using 6,498 flat elements.  The excitation is $x$-polarized with $\hat{k}=-\hat{y}$, $f_0=60$ MHz, and $\fmax=100$ MHz.  In this result, the current is discretized using temporal basis functions of order $\p=0$.  
\begin{figure}[!h]
     \centering
     \includegraphics[width=\linewidth]{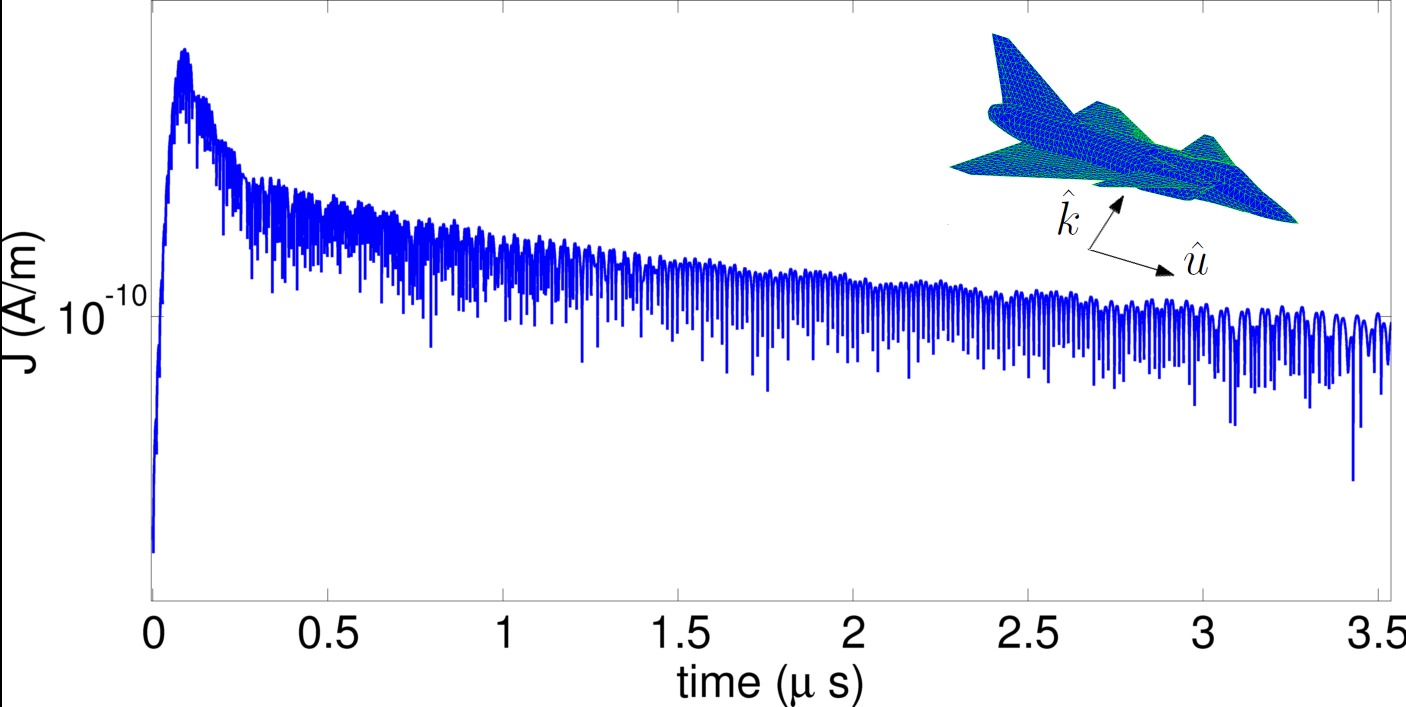}
     \caption{Current coefficient on VFY218}
	\label{vfy_cur}
\end{figure}
This is an extremely challenging problem due to the sharp corners and edges on the geometry, yet the current in figure \ref{vfy_cur}  is stable throughout the simulation (7,000 time steps or 100 transits across the geometry).
\begin{figure}[!h]
     \centering
     \includegraphics[width=\linewidth]{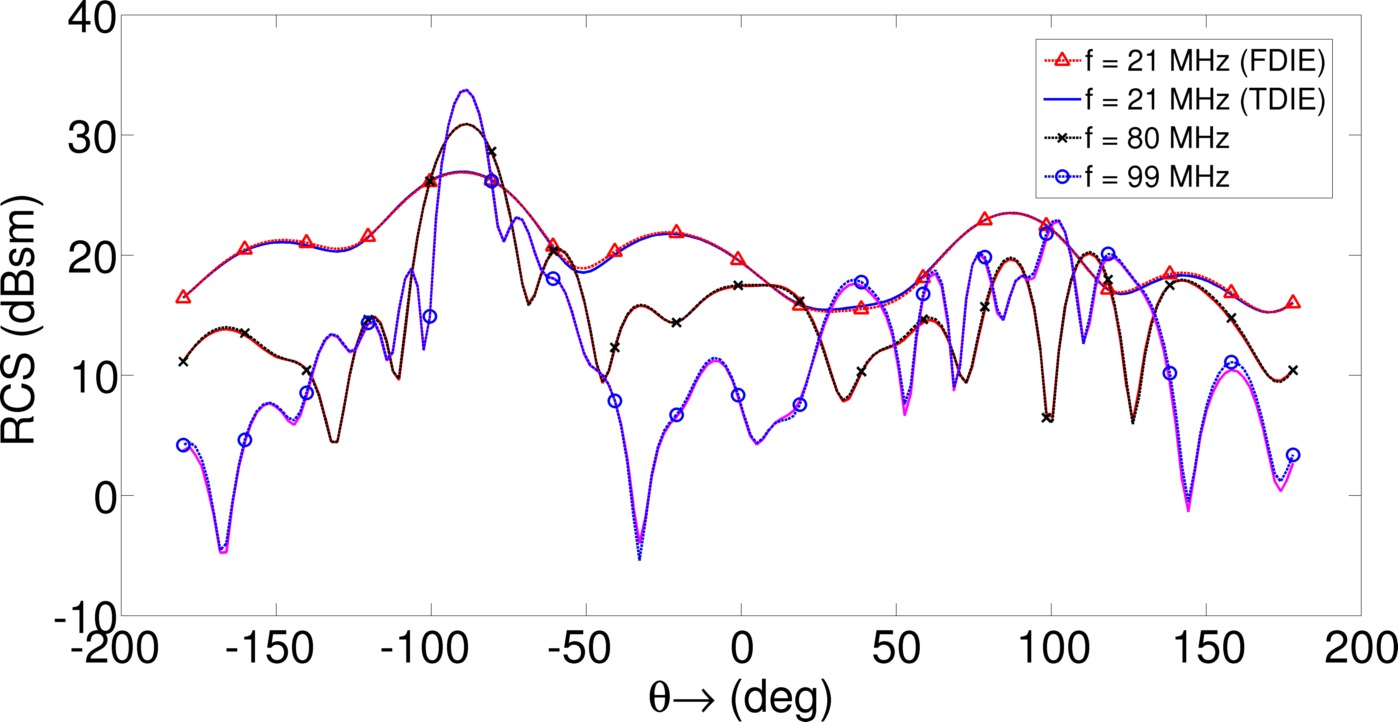}
     \caption{RCS of VFY218}
	\label{vfy_rcs}
\end{figure}
The validity of this result is demonstrated in figure \ref{vfy_rcs}.  The same problem was solved using a frequency domain EFIE solver and the two results are plotted against each other.  Excellent agreement is seen.
\section{Conclusion}
We have presented a higher order (in geometric discretization and temporal basis function order) space-time Galerkin scheme for TDIEs based on a separable expansion in space and time of the retarded potential Green's function.  The expansion yields smooth spatial integrands, which can be evaluated to arbitrary precision through purely numerical means.  This enables the method to be used on higher order tessellations while retaining high accuracy in the matrix elements.  To extend the accuracy of the solver, a higher order temporal basis was used which expands the current using multiple functions within a single time step.  This allows the solver to be used within a space-time Galerkin framework without violating causality.  To validate the method we have presented convergence results for the scattered farfield of a PEC sphere as compared to the Mie series.  To elucidate the stability properties of the solver we have presented stable results of scattering from a variety of geometries, the solution of each being difficult or impossible to stabilize via existing methods.  RCS comparisons have been presented in each example to verify the accuracy of the solution. 
\section{Appendix}
\subsection{Causality of marching system}
Let the surface current be approximated by \eqref{CURRENT_DEF} and let the support of $T_j^l(t)$ lie on the interval $t\in[a,b]$, for all $j>0$, $0\leq j\leq \p$, where $b>a$.  To examine the causality of the marching system we look at the value of 
\begin{equation}
\inprod{T_i^k(t)}{T_j^l(t)\ast_t\delta(t-\Delta)}=\inprod{\delta(t-\Delta)}{T^l(t)\ast_t T^k(-t-(j-i)\dt)}\label{MxEl}
\end{equation}
Causality is maintained if \eqref{MxEl} is zero for $j-i>0$, $\Delta\geq0$, and $t\geq0$.  It can be shown that 
\begin{equation}
\begin{split}
T^l(t)\ast_t &T^k(-t-(j-i)\dt)\neq0\\\text{for}&~(j-i)\dt\in\left((a-b)\dt-t,(b-a)\dt-t\right)
\end{split}
\end{equation}
Considering the extreme case in which $t=0$, the support of this function is seen to include values of $(j-i)>0$ only for case when $b-a>1$.  This shows that the support of $T^l(t)$ must be only one time step.  We note that this also holds true for the important cases of the integral and derivative of $T^l(t)$.  

\subsection{Smoothness of temporal basis functions}
To justify this claim, we examine the time dependent terms in the tested vector potential
\begin{equation}
\inprod{U\left(t\right)}{\dowt^2\delta(t-R/c)\ast T(t)}\label{eq:ApotTemp}
\end{equation}
where $U(t)$ is a compactly supported, discontinuous, locally integrable function and $T(t)$ is compactly supported.  \eqref{eq:ApotTemp} can be written as
\begin{equation}
\int_{-\infty}^{\infty}dt\delta(t-R/c)\int_{-\infty}^{\infty}dt'U(t')\dowtptwo T(t'-t)
\end{equation}
Due to the compactness of $U(t)$ and $T(t)$
\begin{equation}
\int_{-\infty}^{\infty}dt'U(t')\dowtp^2T(t'-t)=-\int_{-\infty}^{\infty}dt'\dowtp U(t')\dowt T(t'-t)
\end{equation}
which is finite if $T(t)$ is at least weakly differentiable to first order, i.e. $T(t)$ is continuous.  For the undifferentiated TD-EFIE, the quantity of interest is
\begin{equation}
\int_{-\infty}^{\infty}dt'U(t')\dowtp T(t'-t)=-\int_{-\infty}^{\infty}dt'\dowtp U(t')T(t'-t)
\end{equation}
which will be finite if $T(t)$ is integrable.

\subsection{Convergence of \eqref{eq:SepExp}}\label{subsec:Nhconv}
The error incurred through truncation \eqref{eq:SepExp} can be determined via passage to the Fourier domain in both $t$ and ${\bf r}$, yielding quantities depending on variables $\omega$ and $\blambda$, respectively.  Given maximum temporal and spatial frequencies of interest, $\omega_{max}$ and $\blambda_{max}$, the bound on this error can be shown to be
\begin{equation}
\begin{split}
 (\ztzs)^\Nh\left[K_1\frac{2\ztzs}{(1-\ztzs)}-K_2(2\Nh+1)\left(\frac{\ztzs}{1-\ztzs}\right)^2\right]
\end{split}
\end{equation}
where
\begin{equation}
z_t=\frac{e\omega_{max}}{k_1(\Nh+3/2)}~~~,~z_s=\frac{e\abs{\blambda_{max}}c}{k_1(\Nh+3/2)}
\end{equation}
and $K_1$ and $K_2$ are some positive constants.  This bound converges rapidly for values of $\ztzs<1$.
\section{Acknowledgement}
The authors would like to acknowledge support under NSF grant CCF1018516 and the DoD SMART Program under grant N00244-09-1-0081.

\bibliographystyle{ieeetr}

\end{document}